\documentclass[letterpaper]{JHEP3}

\usepackage{amsmath,amsthm, amssymb}
\usepackage{epsfig,multicol}

\newcommand{\be}{\begin{equation}}
\newcommand{\ee}{\end{equation}}
\newcommand{\ba}{\begin{array}}
\newcommand{\ea}{\end{array}}
\newcommand{\bea}{\begin{eqnarray}}
\newcommand{\eea}{\end{eqnarray}}
\newcommand{\bma}{\begin{matrix}}
\newcommand{\ema}{\end{matrix}}
\newcommand{\bpm}{\begin{pmatrix}}
\newcommand{\epm}{\end{pmatrix}}

\newcommand{\T}{{\cal{T}}}

\title{Radion effective theory in the detuned Randall-Sundrum model}
\author{Jonathan Bagger {\it and} Michele Redi\\
Department of Physics and
Astronomy, The Johns Hopkins University, \\
3400 North Charles Street, Baltimore, MD 21218, USA\\
E-mail: \email{bagger@jhu.edu}\\
E-mail: \email{redi@pha.jhu.edu}}

\preprint{}

\abstract{We compute the two-derivative low-energy effective
action for the radion in the (supersymmetric) Randall-Sundrum
scenario with detuned brane tensions.  At the classical level, a
potential automatically stabilizes the distance between the
branes. In the supersymmetric case, supersymmetry can be broken
spontaneously by a vacuum expectation value for the fifth
component of the graviphoton.}


\begin{document}

\section{Introduction}

In the original proposal by Randall and Sundrum (RS) \cite{rs},
five-dimensional anti-de Sitter space (AdS$_5$) is truncated by
two branes located at the fixed points of an $S^1/\mathbb{Z}_2$
orbifold.  The branes have opposite tensions, related in magnitude
to the bulk cosmological constant.  In this scenario, the distance
between the branes is a modulus; an additional mechanism is
required to stabilize the radius of the circle.  In contrast, when
the tensions are appropriately detuned, the five-dimensional Einstein's
equations automatically determine the radius \cite{kaloper}.

In this paper, we derive the low-energy effective theory for the
radion in the RS model with detuned brane tensions.  We first
consider the supersymmetric version of the model, constructed in
\cite{bagger2,old}, in which the low-energy dynamics are
controlled by an ${\cal N}=1$ supersymmetric effective lagrangian.
We compute the K\"ahler potential and the superpotential for the
radion superfield.  We then extend these results to the
non-supersymmetric case (where only gravity is present).

This paper is organized as follows.  In section \ref{model} we
review the supersymmetric RS model with detuned brane tensions,
and describe in detail the case where the ground state is AdS$_4$.
In section \ref{noscale} (and appendix A) we determine the general
form of the low-energy effective theory, compatible with the
symmetries of the five-dimensional action.  The low-energy
effective theory turns out to be closely related to no-scale
supergravity. In section \ref{effectivetheory} we use a
Kaluza-Klein reduction to compute the low-energy effective action
for the bosonic fields, valid to two derivatives.  In section
\ref{matching} we match this result to the supersymmetric
effective action and determine all the free parameters.  We also
show that, in contrast to the tuned case, supersymmetry can be
spontaneously broken by a vacuum expectation value (VEV) for the
fifth component of the graviphoton. We summarize the results in
section \ref{summary}. In appendix \ref{c} we discuss corrections
to the matching conditions, while in appendix \ref{compensator} we
present a heuristic derivation of the K\"ahler potential and
superpotential based on the superconformal approach to
supergravity.

\section{Detuned Randall-Sundrum model}
\label{model}

We start by reviewing the supersymmetric RS model with detuned
brane tensions \cite{bagger2}.  The fifth dimension spans the
covering space of an $S^1/\mathbb{Z}_2$ orbifold, parameterized by
the coordinate $\phi$, with  $\phi\in(-\pi,\pi]$.  Three-branes,
with tensions $T_0$ and $T_\pi$, are placed at the orbifold fixed
points, $\phi=0$ and $\phi=\pi$. The five-dimensional action is
given by AdS supergravity, together with brane actions that
describe the tensions of the branes.  The bosonic part of the
action is
\begin{eqnarray}
S_{\rm bulk}&=&-M_5^3\int d^4x\int^\pi_{-\pi}d\phi
\sqrt{-G}\Big(\frac 1 2 {\cal R}-6 k^2+\frac 1 4 F^{MN}F_{MN}+\frac 1
{6\sqrt{6}}
\epsilon^{MNPQR}B_M F_{NP}F_{QR}\Big)\nonumber\\
S_{\rm brane}&=& -T_0 \int d^4x \sqrt{-g_0}-T_\pi \int d^4x
\sqrt{-g_\pi}, \label{action}
\end{eqnarray}
where $g_0$ and $g_\pi$ are the induced metrics on the branes, and
$B_M$ is the graviphoton, a $U(1)$ gauge field required by
supersymmetry.  Supersymmetry restricts the brane tensions to
satisfy the bound $|T_{0,\pi}| \le T$ \cite{bagger2}, where $T$ is
the ``fine-tuned tension,'' related to the five-dimensional Planck
mass by
\begin{equation}
T=6 M_5^3 k.
\end{equation}
When this bound is satisfied, the full bulk-plus-brane theory is
invariant under five-dimensional ${\cal N}=2$ supersymmetry in the
bulk, restricted to four-dimensional ${\cal N}=1$ supersymmetry on
the branes. In this case, the low-energy effective action is
${\cal N}=1$ supersymmetric.

In the original RS model, with tuned brane tensions, the
ground-state metric is flat four-dimensional spacetime, warped
along the fifth dimension.  The Einstein equations are solved for
any distance between the branes; the radius is a modulus of the
compactification.  For detuned brane tensions, four-dimensional
flat space is replaced by AdS$_4$ or dS$_4$.   The AdS$_4$
ground state arises when $|T_{0,\pi}| < T$; the dS$_4$ vacuum
corresponds to $|T_{0,\pi}| > T$.  In the latter case, we exclude
same-sign tensions because they give
a warp factor with a zero between the two branes.  (We do not
consider the other regions for $T_0$
and $T_\pi$ because they do not support parallel brane solutions.)

We first consider AdS$_4$.  The five-dimensional metric is
\begin{equation}
ds^2 = F(\phi)^2\, g_{mn}\, dx^m dx^n+r_0^2\, d\phi^2,
\label{ansatz}
\end{equation}
where $r_0$ is the radius of $S^1$, $F$ is the warp factor, and
$g_{mn}$ is the metric of AdS$_4$ with radius $L$.
The warp factor is determined by the five-dimensional Einstein
equations,
\begin{eqnarray}
F F^{''} - k^2 r_0^2\, F^2  &=& -\frac{2}{ T}\Big[k r_0 T_0 \,F^2
\delta(\phi)
+k r_0 T_\pi \,F^2  \delta(\phi-\pi) \Big] \nonumber \\[2mm]
F^{'2}- k^2 r_0^2\, F^2  &=& -\frac {r_0^2} {L^2}.
\label{bgequation}
\end{eqnarray}
The solution to these equations, defined on the orbifold, is a
combination of positive and negative exponentials,
\begin{equation}
F(\phi)=e^{-k r_0 |\phi |}+\frac 1 {4 k^2 L^2}e^{k r_0 |\phi|}.
\label{warpfactor}
\end{equation}

With our conventions, the radius of the AdS$_4$ metric is related
to the tension of the brane at zero by
\begin{equation}
\frac 1 {4 k^2 L^2}=\frac {T-T_0} {T+T_0}.
\end{equation}
When $T \ne T_0$, the five-dimensional metric is a
parametrization of AdS$_5$ that admits a foliation with AdS$_4$
or dS$_4$ slices of varying curvature along the fifth dimension.
When $|T_{0,\pi}| < T$, the parameter $L$ is real, the
ground state is AdS$_4$, and the five-dimensional theory can be
made supersymmetric.  In contrast, when $|T_{0,\pi}| > T$, the
ground state is dS$_4$.  In that case, certain results can be
obtained by the analytic continuation $L \to i L$.

An intriguing new feature of the detuned scenario is the fact that
the brane tensions fix the radius of the fifth dimension,\footnote{For
simplicity, we restrict the parameters so that $r_0>0$.
In the AdS$_4$ case, this corresponds to $T_0+T_\pi>0$; in
the dS$_4$ case, to $T_0+T_\pi<0$.  A similar analysis holds
when the signs are reversed.}
\begin{equation}
2\pi k r_0= \log \frac {(T+T_0)(T+T_\pi)} {(T-T_0) (T-T_\pi)}.
\label{radius}
\end{equation}
The tensions can be chosen such that the radius is stabilized in
the phenomenologically relevant regime where the warp factor is
large. In the tuned case, when $T_0=-T_\pi=T$, the radius is a
modulus, so $r_0$ is not determined by (\ref{radius}).  Note that
if only one tension is tuned, the critical distance is infinite.

\section{Supersymmetric effective action}
\label{noscale}

In this section we present the most general low-energy effective
action compatible with the symmetries of the supersymmetric
five-dimensional theory.  We will see that the ${\cal N}=1$
supersymmetric effective action is determined up to four free
parameters.

We start by recalling that $G_{mn}$, $G_{55}$ and $B_5$ are even
under the orbifold projection, while $G_{m5}$ and $B_m$ are odd.
Therefore, at energies below the Kaluza-Klein (KK) scale, the
bosonic effective action includes fluctuations of the
four-dimensional metric $g_{mn}$, together with the light modes of
$G_{55}$ and $B_5$. As we will see, the scalar associated with
$G_{55}$ is correctly identified as the proper distance between
the branes. The other scalar is the zero mode of $B_5$, since the
VEV of $B_5$ is a modulus of the compactification.  The fields
$G_{m5}$ and $B_m$ do not give rise to any light modes, but their
tadpoles can still contribute to the low-energy effective action.

In the supersymmetric effective theory, the two scalars join with
the fifth component of the gravitino to form a chiral
supermultiplet, given by
\begin{equation}
D(E,0)\oplus D(E+1,0)\oplus D(E+\frac 1 2,\frac 1 2),
\end{equation}
where $D(E,s)$ denotes the AdS$_4$ representation labeled by the
Casimir $E$ and the spin $s$. The masses of the particles are
related to the eigenvalues $E$,
\begin{equation}
m_0^2(E)=\frac {E(E-3)} {L^2}, \qquad\qquad m_{1/2}(E)=\frac
{E-3/2} {L}. \label{reps}
\end{equation}
Since the value of $B_5$ is a modulus, the corresponding zero mode
must be massless.  From (\ref{reps}), there are only two possible
masses for its scalar partner,
\begin{equation}
m_r^2=- \frac 2 {L^2} \qquad {\rm or} \qquad m_r^2=\frac 4 {L^2}.
\end{equation}
As we will see in the next section, the KK reduction sets $m_r^2=
4 / {L^2}$.  This fixes the mass of the fermionic superpartner to
be $m_{1/2}=2/L$.

The effective action is ${\cal N}=1$ supersymmetric, so it is
determined by a K\"ahler potential $K$ and a superpotential $P$.
The bosonic part of the action (setting $M_4=1$) takes the form
\begin{equation}
S_{\rm eff}=- \int d^4x \sqrt{-g} \Big[\frac{1}{2} R+K_{\T
\overline{\T}}g^{mn}\partial_m \T \partial_n \overline{\T} + e^{K}
(K^{\T\overline{\T}}D_\T P {D_{\overline \T} \overline{P}}- 3
P\overline{P})\Big],
\end{equation}
where $\T$ is the lowest component of the radion superfield, and
$D_\T P=\partial_\T P+ K_\T P$ is the covariant derivative of the
superpotential. The imaginary component of $\T$ can be readily
identified with the zero mode of $B_5$. In the next section, we
will use the KK reduction to prove that the real part of $\T$ is
the proper distance between the branes.

To find $K$ and $P$, we recall that the \emph{bosonic} part of the
five-dimensional action is invariant under the shift $B_5 \to B_5+
{\rm constant}$. This follows from the fact that $B_5$ is
derivatively coupled in (\ref{action}), except for an irrelevant
$F\tilde{F}$ term. It implies that the K\"ahler potential is a
function of $\T+\overline{\T}$ (up to a K\"ahler transformation).
It also implies that the scalar potential,
\begin{equation}
V(\T,\overline{\T}) =e^{K}(K^{\T \overline{\T}}D_\T P
{D_{\overline \T} \overline{P}}-3 P\overline{P}),
\end{equation}
cannot depend on the zero mode of $B_5$.  This is an extremely
restrictive condition because the superpotential, being a
holomorphic function of $\T$, depends explicitly on the zero mode.
In appendix A we show that, in an AdS$_4$ ground state, the most
general solution to this condition is\footnote{This holds for
$m_r^2=4/L^2$.}
\begin{eqnarray}
K(\T+\overline{\T})&=&-3 \log[1-c^2\,
e^{-a(\T+\overline{\T})}]\nonumber\\[2mm]
P(\T)&=&p_1+p_2\, e^{-3 a  \T}, \label{kahler}
\end{eqnarray}
where $p_1$, $p_2 \in \mathbb{C}$ and $a$, $c \in \mathbb{R}$ are
undetermined constants.

With an obvious change of variables, we can cast our result in a
more familiar form,
\begin{eqnarray}
K(z,\overline{z})&=&-3 \log[1-z\overline{z}] \nonumber \\[2mm]
P(z)&=& p'_1+ p'_2 z^3.
\end{eqnarray}
In this parametrization, the K\"ahler potential is that of
no-scale supergravity \cite{noscale}.  In particular, the scalars
are the coordinates of the manifold $SU(1,1)/U(1)$.  The
superpotential is the generalization to AdS space of the
superpotential of ordinary no-scale supergravity. The
superpotential breaks explicitly the $SU(1,1)$ symmetry of the
kinetic term. However, the bosonic action preserves a $U(1)$
subgroup that corresponds to multiplying $z$ by a phase.

\section{Bosonic reduction}
\label{effectivetheory}

As we have seen, the low-energy effective theory depends on four
constants. To determine the constants, we match the effective
action derived from (\ref{kahler}) to the bosonic reduction of the
five-dimensional theory.

\subsection{Radion reduction}

To compute the low-energy effective action for the radion, we
consider the following ansatz for the metric,
\begin{equation}
ds^2=\Big( F(\phi)^2 + A(x) \Big) g_{mn}(x)\,dx^m dx^n + r_0^2\,
\Bigg( \frac {F(\phi)^2} {F(\phi)^2+A(x)} \Bigg)^2 d\phi^2.
\label{radionansatz}
\end{equation}
This ansatz is a generalization of the one used in \cite{bagger}.
The field $A$ parameterizes fluctuations in the distance between
the branes; the field $g_{mn}$ describes the four-dimensional
graviton.  For $A=0$, the branes are at the critical distance and
the five-dimensional equations of motion require $g_{mn}$ to
satisfy the four-dimensional Einstein equations with a
cosmological term,
\begin{equation}
R_{mn}-\frac 3 {L^2} g_{mn}=0.
\end{equation}
The static solution to this equation is AdS$_4$ with radius of
curvature $L$. For $A \ne 0$ there is no static solution,
reflecting the fact that a potential is generated for $A$.

The virtues of this ansatz are three-fold.  First, it ensures that
there are no tadpoles associated with the $G_{m 5}$ components of
the metric (see section \ref{tadpoles}).  Second, the low-energy
effective action is automatically in Einstein frame:  to any order
in the fields, $A$ does not mix with the four-dimensional
graviton. Third, the ansatz respects the boundary conditions that
follow from the brane actions,
\begin{equation}
\omega_{m a\hat{5}}\Big|_{\phi\sim 0}=\epsilon(\phi) \frac {T_0}
{T} e_{m a}, \qquad\qquad \omega_{m a\hat{5}}\Big|_{\phi\sim
\pi}=-\epsilon(\phi) \frac {T_{\pi}} {T} e_{m a},
\end{equation}
where $\omega_{M A B}$ is the spin connection and $e_{m a}$ is the
four-dimensional vierbein. As shown in \cite{bagger2}, these
conditions are also required by local supersymmetry in the
five-dimensional theory.

A naive effective action can be found by substituting the ansatz
(\ref{radionansatz}) into (\ref{action}) and integrating over the
fifth dimension.  This gives
\begin{equation}
S_4=-\int d^4 x \sqrt{-g}\Big[\frac {M_4^2} 2  R-\frac {3 M_4^2}
{L^2} +\alpha(A) g^{mn}\partial_m A\partial_n A+(T_0+T_\pi)
A^2\Big], \label{aeffective}
\end{equation}
where
\begin{equation}
\alpha(A)=\frac {T\,r_0\,} {8\,k\,} \int_{-\pi}^{\pi}\frac
{F(\phi)^2} {\big(F(\phi)^2+A\big)^2}d\phi
\end{equation}
and we have introduced the four-dimensional Planck mass
\begin{equation}
M_4^2=\frac {T\,r_0\,} {6\,k\,}  \int_{-\pi}^{\pi} F(\phi)^2
d\phi.
\end{equation}
As advertised, the ansatz gives an effective action in the
Einstein frame.  Note that the effective action contains a
potential for $A$, enforcing the condition $A=0$ in the
ground state.

It is convenient and more physical to write the effective action
in terms of the proper distance $\pi r$ between the branes. The
radion field $r$ is related to the fluctuation $A$ as follows,
\begin{equation}
r(x)=\frac 1 {2\pi} \int_{-\pi}^{\pi} \sqrt{G_{55}} d\phi=\frac
{r_0} {2\pi} \int_{-\pi}^{\pi} \frac {F(\phi)^2} {F(\phi)^2+A(x)}
d\phi. \label{radion}
\end{equation}
The function $\alpha$ is then simply
\begin{equation}
\alpha(A)=-\frac {\pi } {4}\, \frac{T}{k}\, \frac {\partial r}
{\partial A}.
\end{equation}
With the change of variables (\ref{radion}), the effective action
takes the form
\begin{equation}
S_4=-\int d^4x \sqrt{-g}\Big[\frac {M_4^2} 2 R-\frac {\pi } {4}\,
\frac{T}{k}\, \Big(\frac {\partial A}{\partial r}\Big) g^{mn}
\partial_m r\partial_n r+V(A(r))\Big], \label{radioneffective}
\end{equation}
where
\begin{equation}
V(A(r))= -\frac {3 M_4^2} {L^2} +(T_0+T_\pi) A^2.
\end{equation}
In these expressions, all quantities are expressed as functions of
$r$.  The potential has a minimum at $r=r_0$, with
four-dimensional cosmological constant
\begin{equation}
\Lambda_4=V\Big|_{r=r_0}=-\frac {3 M_4^2} {L^2}.
\end{equation}

\subsection{Graviphoton reduction}
\label{graviphoton}

The graviphoton reduction proceeds as in \cite{bagger,sundrum}.
The five-dimensional equations of motion for the graviphoton are
\begin{equation}
\partial_M\big(\sqrt{-G}G^{MN}G^{PQ}F_{NQ}\big)=0,
\label{5dgraviphoton}
\end{equation}
where we have dropped the contribution of the Chern-Simons term
since it does not contribute to the two-derivative effective
action.  In the metric background, (\ref{5dgraviphoton}) becomes
\begin{eqnarray}
&&\partial_m\Big(\sqrt{-g} F^{-2}\big(F^2+A \big)^2 g^{mn} F_{5n}\Big)=0 \nonumber\\
&&\partial_5\Big(\sqrt{-g}F^{-2}\big(F^2+A\big)^2
g^{mn}F_{5n}\big)+
\partial_p\Big(\frac {r_0^2 F^2} {F^2+A}\sqrt{-g}g^{pq}g^{mn}F_{q n}\Big)=0,
\label{beqn}
\end{eqnarray}
We identify the graviphoton zero mode $b$ with the Aharonov-Bohm
phase of $B_M$ around the fifth-dimension
\begin{equation}
b(x)=\frac 1 {\sqrt{6}\pi}\int_{-\pi}^{\pi}  B_5 d\phi, \label{b}
\end{equation}
where the normalization is chosen for later convenience. The field
$b$ is gauge invariant for gauge transformations that are periodic
on the circle.

Because of the parity assignments, the components $B_m$ do not
produce light modes. However, as in \cite{bagger,sundrum}, they
contribute to the effective action through the tadpole $\partial_5
B_m \partial^m B_5$.  From equations (\ref{beqn}), we learn that
\begin{equation}
F_{m5}(x,\phi)=c_m(x) \frac {F(\phi)^2}
{\big(F(\phi)^2+A(x)\big)^2}~~~~~~~~~~~~~~~~~~F_{mn}=0,
\end{equation}
to lowest order in the derivative expansion. The unknown function
$c_m$ can be determined by integrating over the circle.  Since all
the fields are periodic, we find
\begin{equation}
\partial_m b(x)= \frac{c_m(x)}{\sqrt{6} \pi} \int_{-\pi}^\pi
\frac {F(\phi)^2} {\big(F(\phi)^2+A(x) \big)^2} d\phi,
\end{equation}
from which it follows that
\begin{equation}
F_{m5}=\sqrt{6} \pi \Big(\int_\pi^\pi  \frac {F(\phi)^2}
{\big(F(\phi)^2+A \big)^2} d\phi\Big)^{-1}\frac {F(\phi)^2}
{\big(F(\phi)^2+A \big)^2}~
\partial_m b.
\end{equation}
Notice that the constraint $dF=0$ implies that the field strength
$F_{mn}$ is already of two-derivative order, so it does not
contribute to the two-derivative effective action.

The effective action for the graviphoton is obtained by
substituting $F_{5m}$ into (\ref{action}).  This gives
\begin{eqnarray}
S_4&=&-\frac {\pi^2 } {2}\, \frac {T} {k r_0}\int d^4x \sqrt{-g}
\Big(\int \frac {F(\phi)^2}
{\big(F(\phi)^2+A\big)^2}d\phi\Big)^{-1}g^{mn}\partial_m b
\partial_n b \nonumber \\
&=&\frac {\pi } {4}\, \frac {T} {k } \int d^4x \sqrt{-g}
\Big(\frac {\partial A} {\partial r}\Big) g^{mn}\partial_m
b\partial_n b. \label{beff}
\end{eqnarray}
The complete bosonic effective action is obtained by combining
(\ref{radioneffective}) with (\ref{beff}),
\begin{equation}
S_4=-\int d^4x \sqrt{-g}\Big[\frac {M_4^2} 2 R-\frac {\pi } {4}\,
\frac {T} {k } \Big(\frac {\partial A}{\partial r}\Big) g^{mn}
\big(\partial_m r\partial_n r+\partial_m b\partial_n
b\big)+V(A(r))\Big]. \label{effective}
\end{equation}
The kinetic terms appear in K\"ahler form (in fact for any choice
of $F$). This shows that the proper distance $r$ is the correct
supersymmetric variable.

\subsection{Gravitational tadpoles}
\label{tadpoles}

At the classical level, the effective action derived above
receives corrections associated with integrating out the heavy KK
modes. In flat space, the effective action can be organized in a
derivative expansion, in which higher-derivative terms are
suppressed by appropriate powers of the cut-off.  At energies much
lower than the cut-off, the higher-derivative terms are
negligible. The situation is more subtle in AdS space because a
new scale appears, the curvature $1/L^2$.  Higher-derivative
operators such as $R^n$ do not vanish in the vacuum, but give
contributions to the action of order $1/L^{2n}$. Therefore,
consistency requires that the two-derivative effective theory also
be expanded to first order in $1/L^2$.

To see how this works, let us compute the five-dimensional
equations of motion.  For simplicity, we work with the variable
$A$.  Using the ansatz (\ref{radionansatz}), the purely
gravitational Einstein equations read
\begin{eqnarray}
R_{mn}&=&\frac 3 {L^2}g_{mn} -\frac 3 2 \frac 1
{\big(F(\phi)^2+A\big)^2}
\partial_m A\partial_n A-\frac 3 {L^2} \frac {A^2}
{F(\phi)^4}g_{mn} \nonumber\\[2mm]
\square A&=&\frac 4 {L^2} A+ \frac 1 2 \frac {\partial^m
A\,\partial_m A} {F(\phi)^2+A} +\frac 4 {L^2} \frac {A^2}
{F(\phi)^2}. \label{consistency}
\end{eqnarray}
(We set $B_M=0$ because the graviphoton is not relevant for the
present discussion.)  The $(m 5)$ equations are identically zero.

Equations (\ref{consistency}) show that with our ansatz, the
five-dimensional equations of motion are inconsistent.  The
left-hand side of each equation is a function of the $x^{m}$ only,
while the right-hand side depends on the coordinate $\phi$.  Note,
however, that to {\it linear} order in $A$, the five-dimensional
equations are consistent; they are satisfied point-by-point in the
extra dimension.  In particular, the second equation determines
the mass of $A$,
\begin{equation}
m_A^2=\frac 4 {L^2}.
\end{equation}
This is also the mass of the radion field, since $r$ is related to
$A$ by a change of variables.\footnote{The same result was
previously found in \cite{chacko}. In fact, our ansatz coincides
with theirs to linear order.} As we have seen, this value of the
mass is required by supersymmetry.

The inconsistency of the five-dimensional equations reflects the
fact that the Kaluza-Klein gravitons mix with the radion through
tadpoles, so the heavy KK fields cannot consistently be set to
zero.  In fact, eqs.\ (\ref{consistency}) are inconsistent at
two-field order.  This implies that the five-dimensional action
contains terms with one heavy field coupled to two light fields.
Schematically, the action for these fields takes the form,
\begin{equation}
{\cal L}_{tadpoles}= A^2 H+(\partial H)^2+M^2 H^2+\dots,
\label{tadpolesaction}
\end{equation}
where $H$ stands for a generic heavy field.  Note, though, that
the $A^2 H$ couplings contain at least two derivatives or two
powers of $1/L$.  Therefore, integrating out the heavy fields
modifies the effective action for the light fields at order
$\partial^4$, $\partial^2/L^2$, or $1/L^4$. As a consequence, the
effective action (\ref{effective}) obtained in the previous
section should be trusted only up to leading order in $\partial^2$
and $1/L^2$.  To that order, eq.~(\ref{effective}) becomes
\begin{equation}
S_{\rm eff}=- \int d^4x \Big[ \frac{\hat{M}_4^2}{2}R  + 3
 \pi^2 k^2 \hat{M}_4^2\, \frac{e^{- \pi k (\T+\overline{\T})}} {(1-
e^{-\pi k(\T+\overline{\T})})^2}
 g^{mn}\partial_m \T \partial_n \overline{\T}+ V(\T,\overline{\T}) \Big],
\label{susyeffective2}
\end{equation}
where
\begin{equation}
V(\T,\overline{\T}) = -\frac {3\hat{M}_4^2\,(1- e^{-2 \pi k
r_0})}{L^2}\left[ \frac{1-e^{-2 \pi k (\T+
\overline{\T}-r_0)}}{(1-e^{-\pi k  (\T+ \overline{\T})})^2}\right],
\end{equation}
$\T=r+ib$ and $\hat{M}^2_4 = M^2_4 = (T/6 k^2)(1 -
e^{-2\pi k r_0})$, to leading order in $1/L^2$.

\section{Results}
\label{matching}
\subsection{Supersymmetric action}

In section 3 we showed that the most general supersymmetric
action, compatible with the symmetries of the five-dimensional
theory, is specified by
\begin{eqnarray}
K(\T+\overline{\T})&=&-3 M^2_4 \log[1-c^2\,
e^{-a(\T+\overline{\T})}]\nonumber\\[2mm]
P(\T)&=&p_1+p_2\, e^{-3 a \T},
\end{eqnarray}
where $\T=r+i b$ and we have restored the four-dimensional Planck
mass, $M_4$.

The parameters $a$, $c$, $p_1$ and $p_2$ must be determined by
matching with the KK reduction of the five-dimensional theory.  In
particular, $p_1$ and $p_2$ are set by matching the cosmological
constant of the AdS$_4$ ground state and by requiring that
the minimum of the potential be at $r=r_0$. This gives
\begin{equation}
p_1 =  \frac{ M_4^2 \sqrt{1-c^2 e^{-2 a r_0}}} {L}, \qquad\qquad
p_2 = -\frac{c^2 e^{a r_0} M^2_4 \sqrt{1-c^2 e^{-2 a r_0}}}{
L}~e^{i\beta},
\end{equation}
where, for the moment, the phase $\beta$ is free.

These results fix the bosonic part of the low-energy effective
action in terms of $a$ and $c$.  Using (5.1) and (5.2), we find
\begin{equation}
S_{\rm eff}=- \int d^4x \Big[ \frac{M_4^2}{2}R  + 3 a^2\, c^2
M_4^2 \, \frac{e^{-a (\T+\overline{\T})}} {(1- c^2\, e^{-a
(\T+\overline{\T})})^2}
 g^{mn}\partial_m \T \partial_n \overline{\T}+ V(\T,\overline{\T}) \Big],
\label{susyeffective}
\end{equation}
where the scalar potential is
\begin{equation}
V(\T,\overline{\T}) = -\frac {3\,M_4^2\,(1-c^2 e^{-2 a
r_0})}{L^2}\left[ \frac{1-c^2\,e^{-2a(\T+
\overline{\T}-r_0)}}{(1-c^2\,e^{-a  (\T+
\overline{\T})})^2}\right]. \label{potential}
\end{equation}
As expected, the potential is independent of $b$ (and the phase
$\beta$).

The parameters $a$ and $c$ can be found by matching
(\ref{susyeffective}) with the effective action
(\ref{susyeffective2}). This gives
\begin{equation}
a\ =\  k \pi, \qquad \qquad c \ =\  1, \label{parameters}
\end{equation}
along with $M^2_4 = \hat{M}^2_4$, up to corrections of order
$1/L^2$ (see appendix B). The K\"ahler potential is
then\footnote{These results agree with \cite{brax}, where the
authors considered the case $T_0=-T_\pi$. In this
limit, the critical distance between the branes is zero.}
\begin{equation}
K(\T+\overline{\T})= -3  \hat{M}^2_4 \log[1- e^{-\pi
k(\T+\overline{\T})}],
\end{equation}
while the superpotential is
\begin{equation}
P(\T)=\frac {k \hat{M}^3_4}{L} \sqrt{ \frac{6}{ T}}  \left(1 -e^{i
\beta} e^{\pi k r_0}  e^{-3 \pi k \T}\right).
\label{superpotential}
\end{equation}
The superpotential can be understood as the sum of two constant
superpotentials, one localized at each of the orbifold fixed
points. The radion dependence results from the warping of the
metric in the extra dimension. In appendix C we show that this
result can be explained in terms of the conformal compensator
approach to supergravity.

\subsection{Continuation to dS$_4$}
\label{ds4}

Since it is consistent with the equations of motion to set
$B_M=0$, the effective theory derived above gives the effective
action even in the non-supersymmetric case, where the only field
is the metric $G_{MN}$.

The same approach can also be used when the four-dimensional
ground state is dS$_4$ (see also \cite{ito}).  In this case the
absolute values of the brane tensions both exceed the tuned value.
As explained in section 2, we take the tensions to have opposite
signs. Furthermore, since the five-dimensional theory cannot be made
supersymmetric \cite{bagger2}, we only consider pure gravity.

To compute the effective theory for the radion, we proceed as in
section \ref{effectivetheory}.  We use the ansatz
(\ref{radionansatz}) to derive the effective action, up to two
derivatives and leading order in $1/L^2$. From the linearized
analysis, we find that the radion has a tachyonic mass
\cite{chacko}, $m_R^2=-4/L^2$, which implies that the dS$_4$
solution is unstable.

We can now exploit the fact that the five-dimensional equations of
motion are identical to the AdS$_4$ case, with $L$ replaced by $i
L$. Therefore, even though the theory is not supersymmetric, the
low-energy effective action can be obtained from the
supersymmetric result. The kinetic term remains the same, while
the potential changes sign,
\begin{equation}
V(\T,\overline{\T}) = \frac {3 \hat{M}^2_4 (1-e^{-2 \pi k
r_0})}{L^2} \left[ \frac{1-\,e^{-2\pi k(\T+ \overline{\T} -
r_0)}}{(1-\,e^{-\pi k (\T+ \overline{\T})})^2}\right].
\end{equation}
Given the K\"ahler potential (\ref{kahler}), one can show that no
superpotential can give rise to this scalar potential, in accord
with the fact that the original theory is not supersymmetric.

\subsection{Supersymmetry breaking}
\label{susybreaking}

The bosonic reduction completely determines the full
supersymmetric effective action, up to the single phase $\beta$ in
the superpotential (\ref{superpotential}). Without loss of
generality, the value of $\beta$ can be fixed by demanding that
supersymmetry be unbroken when $b=0$. In \cite{bagger2}, it was
shown that one can always choose fermionic brane actions in such a
way that this is true.  Other choices of brane action correspond
to shifting the origin of $B_5$ \cite{us}.

In the effective theory, unbroken supersymmetry requires that the
covariant derivative of the superpotential vanishes, when
evaluated at the minimum of the potential.  With the above
K\"ahler potential and superpotential, we find
\begin{equation}
D_\T P\big|_{r=r_0} \sim 1-e^{i \beta} e^{i 3  \pi k b},
\end{equation}
which implies that we must set $\beta=0$. With $\beta=0$,
supersymmetry is then spontaneously broken when $\langle b \rangle
\ne 2 n/ (3k)$, for $n$ integer. This mechanism of supersymmetry
breaking is the generalization to AdS$_4$ of the no-scale models
with constant superpotentials.
In the limit where the tensions are tuned, $T_0=-T_\pi=T$,
supersymmetry cannot be broken because the superpotential
(\ref{superpotential}) vanishes identically.

In ref.\ \cite{us}, by studying the five-dimensional Killing
spinor equations, we found that a non-zero VEV of $b$ breaks
supersymmetry.  This corresponds to a non-trivial Wilson line for
the graviphoton around the fifth dimension. The supersymmetry
breaking vanishes in the tuned limit because the Killing spinor
equations can always be satisfied in that case.  In  \cite{us} we
also found that $b$ is a periodic variable.  The periodicity
matches exactly the one derived here.

\section{Summary}
\label{summary}

In this paper we computed the low-energy effective theory for the
radion in the supersymmetric Randall-Sundrum scenario with detuned
brane tensions, extending the results of \cite{bagger, sundrum}.
The form of the effective action is determined by ${\cal N}=1$
supersymmetry and by the symmetries of the bosonic
five-dimensional theory. The K\"ahler potential and the
superpotential for the radion chiral superfield (to leading order
in $1/L^2$) are given by
\begin{eqnarray}
K(\T,\overline{\T})&=&-3 \log\Big[1-e^{- \pi
k(\T+\overline{\T})}\Big] \nonumber
\\[2mm]
P(\T)&=&\frac {k}{L} \sqrt{ \frac{6}{ T}} \left(1 - e^{\pi k r_0}
e^{-3 \pi k \T}\right), \label{summary2}
\end{eqnarray}
where $T=r+i b$. The ground state is AdS$_4$; the potential
stabilizes the distance between the branes, while $b$ remains a
modulus of the compactification.  This supersymmetric model is a
generalization to anti de-Sitter space of the no-scale
supergravity models of flat space.  Supersymmetry can be broken
spontaneously by a VEV for the fifth component of the graviphoton.

The purely gravitational case is obtained setting graviphoton and
the fermionic degrees of freedom to zero.  The action can also be
continued to dS$_4$, even though the five-dimensional theory is
not supersymmetric. The theory is unstable in this case.

\acknowledgments

The authors would like to acknowledge helpful conversations with
K.~Agashe, P.~Fox, D.~E.~Kaplan, and especially R.~Sundrum. M.~R.
would like to thank J.~Santiago for a very insightful discussion.
J.B. would also like to thank the Aspen Center for Physics, where
part of this work was done.  This work was supported in part by
the U.S.\ National Science Foundation, grant NSF-PHY-9970781.

\appendix

\section{AdS no-scale supergravity}
\label{appendixa}

In this appendix we derive the K\"ahler potential and
superpotential presented in section 3.  The scalar potential for
the complex radion field $\T$ is given by\footnote{In principle,
the real part of $\T$ is a generic function of $r$. 
The following derivation relies only on the fact that the
imaginary part of $\T$ is proportional to the zero mode of
$B_5$.}
\begin{equation}
V(\T,\overline{\T}) =e^{K}(K^{\T \overline{\T}}D_\T P
{D_{\overline \T} \overline{P}}-3 P\overline{P}).
\label{noscaleappendix}
\end{equation}
In this expression, the K\"ahler potential $K$ is a function of
$\T + \overline{\T}$, while the superpotential $P$ is an analytic
function of $\T$.

In what follows we determine $K$ and $P$ so that $V$ is a function
of the real part of $\T$ only. We start by expanding in a power
series about the vacuum, which by a shift we take to be at $\T=0$,
\begin{eqnarray}
K(\T,\overline{\T})&=& \sum_{n=0}^\infty k_n\left(\frac
{\T+\overline{\T}} 2\right)^n\nonumber \\
P(\T)&=&\frac 1 L \sum_{n=0}^\infty p_n \T^n,
\label{noscaleexpansion}
\end{eqnarray}
where $k_n \in \mathbb{R}$ and $p_n \in \mathbb{C}$. We use a
residual K\"ahler transformation to set $k_0$ and $k_1$ to zero;
we take $k_2=2$ to fix the normalization of the kinetic term. We
also set $p_1=0$ so that supersymmetry is unbroken when $\T=0$.

We now substitute (\ref{noscaleexpansion}) into
(\ref{noscaleappendix}) and expand in powers of $x = \rm{Re}\,\T$
and $y = \rm{Im}\,\T$.  To lowest order in the expansion, we must set
$p_0=1$ to ensure that the ground state corresponds to AdS space
of radius $L$. To next order, we find two possibilities, $p_2=0$
and $p_2 =-3/2$.  We choose the latter, which gives the correct
radion mass, $m_x^2=4/L^2$.

Further conditions are found by requiring that terms proportional
to $x^n y^k$ vanish, for $k \ge 1$. The $x^n y$ terms fix the
$p_n$ to be real (up to an irrelevant overall phase).  The other
terms iteratively determine the coefficients $k_n$ and $p_n$ in
terms of $k_3$ for $n \ge 3$. To see how this works, note that at
order $n$, the parameters $(p_n,k_n)$ appear for the first time.
These parameters are fixed by the $x^{n-k}y^k$ terms, for $k>1$.
Explicitly, from the $x^{n-2}y^2$, $x^{n-4} y^4$ terms, we find
\begin{eqnarray}
n(n-1)((4n-20)\,p_n-9 k_n)&=& g_{n-1}(p_{n-1},k_{n-1})\nonumber \\
n(n-1)(n-2)(n-3)(n-13)\, p_n &=& h_{n-1}(p_{n-1},k_{n-1}),
\end{eqnarray}
where the functions $g_n$ and $h_n$ are polynomials that
depend only on the parameters $(p_n,k_n)$ and lower.  (The
equations associated with other powers of $y$ are redundant.)
The above equations determine $p_n$ and $k_n$ for any $n$
except $n=13$.  For $n=13$, the equation associated with
$x^{n-6} y^6$ supplies the missing relation.

This completes the proof that the K\"ahler potential and
superpotential are determined by two physical parameters, $L$ and
$k_3$ (which is related to the three-point function). By a field
redefinition, $K$ and $P$ can be cast in the form presented in
section 3,
\begin{eqnarray}
K(\T+\overline{\T})&=&-3 \log[1-c^2\,
e^{-a(\T+\overline{\T})}]\nonumber\\[2mm]
P(\T)&=&q_1+q_2\, e^{-3 a  \T},
\label{a4}
\end{eqnarray}
with $q_1$, $q_2 \in \mathbb{C}$.  The six parameters
in (\ref{a4})
are $L$ and $k_3$, the scale of $\T$, a shift of $x$ and
$y$, and an overall (irrelevant) phase of $P$.

\section{Corrections to $K$ and $P$}
\label{c}

In this appendix we present evidence that the effective action
(\ref{effective}) is correct to cubic order in the fluctuations of
the radion, and all orders in $1/L^2$.  If true, this would allow
us to compute the K\"ahler potential and superpotential to all
orders in $1/L^2$.  The $1/L^2$ corrections are necessary when
treating the higher-derivative terms in the effective action.

We start by considering the action (\ref{effective}). The kinetic
term for the radion is given by
\begin{equation}
S_{\rm kin}=\frac {\pi} {4}\,\frac {T} {k} \int d^4x\sqrt{-g}\,
\Big(\frac {\partial A}{\partial r}\Big) g^{mn} \partial_m r
\partial_n r. \label{radionkinetica}
\end{equation}
From (\ref{consistency}), we see that the five-dimensional
equations of motion are consistent to linear order in $A$, and all
orders in $1/L^2$.  Therefore the heavy-field tadpoles modify the
effective action, starting at fourth order in $A$.

To write the action explicitly, we need to express the derivative
$\partial A/\partial r$ as function of $r$.  Therefore we must
invert the relation
\begin{equation}
r(A) =\frac {r_0} {2\pi}\int_{-\pi}^{\pi} \frac {F(\phi)^2}
{F(\phi)^2+A} d\phi. \label{radiona}
\end{equation}
Since tadpoles affect the action at fourth order, it is sufficient
to compute $A$ to quadratic order in $r$. Expanding
(\ref{radiona}) in powers of $A$, we find
\begin{equation}
r=r_0\big(1-I_2 A+I_4 A^2+\dots\big), \label{expansiona}
\end{equation}
where we have introduced the integrals
\begin{equation}
I_n=\frac 1 {2\pi}\int_{-\pi}^\pi F(\phi)^{-n} d\phi.
\end{equation}
Equation (\ref{expansiona}) is readily inverted,
\begin{equation}
A=-\frac 1 {I_2} \left(\frac {r-r_0} {r_0}\right)+\frac {I_4}
{I_2^3} \left(\frac {r-r_0} {r_0}\right)^2+\dots
\end{equation}
Using this result, we find that the kinetic term
(\ref{radionkinetica}) becomes
\begin{equation}
S_{\rm kin}=-\frac {\pi} {4}\,\frac {T} {kr_0} \int
d^4x\sqrt{-g}\, \Big[\frac 1 {I_2}-\frac {2 I_4}
{I_2^3}\left(\frac {r-r_0} {r_0}\right)+\dots\Big]\, g^{mn}
\partial_m r \partial_n r. \label{skin}
\end{equation}

The potential term in (\ref{effective}) can also be written in
terms of $r$,
\begin{equation}
V(r) =\frac {\pi r_0} {L^2} \frac {T} {k} I_2\, A^2 = \frac {\pi
r_0} {L^2} \frac {T} {k} \Big[ \frac{(r-r_0)^2}{I_2 r_0^2}-2 \frac
{I_4} {I_2^3}\frac{(r-r_0)^3}{r_0^3}+\dots\Big].
\end{equation}
The coefficient of the cubic term is precisely the one required by
supersymmetry, proving that the effective action obtained by the
Kaluza-Klein reduction is supersymmetric to cubic order in fields,
to {\it any} order in $1/L^2$.  Moreover, expanding the action
(\ref{effective}) to fourth order, one can check that it is not
supersymmetric to that order.

These results suggest that the effective action for the matter
fields is correct to all orders in $1/L^2$ (and cubic order in the
fields). Then, by matching with the supersymmetric action, we can
compute the corrections to the K\"ahler potential and the
superpotential, to all orders in $1/L^2$. The periodicity of $b$
matches that of the five-dimensional theory \cite{us}, so the
relation $a= k \pi$ is exact. However, the identification $c=1$
receives corrections at order $1/L^2$. Matching the cubic action
for $r$ with the corresponding terms in the supersymmetric
effective theory, we find
\begin{equation}
\frac {1-c^2\, e^{-2\pi k r_0}}{1+c^2\, e^{-2\pi k r_0}}= \pi k
r_0 \frac{I_2^2} {I_4}.
\end{equation}
Note that in this case, we do not match the Planck mass because it
is corrected by the higher-derivative operators proportional to
$R^n$.

\section{Conformal compensator approach}
\label{compensator}

The K\"ahler potential and the superpotential found in this paper
have a beautiful explanation in terms of the conformal compensator
approach to supergravity (see \cite{sundrum}). In this formalism
\cite{compensator}, one decouples gravity and retains one of the
supergravity auxiliary fields. The superspace effective lagrangian
for the radion is given by
\begin{equation}
L =\int d^4\theta \overline{\Sigma} \Sigma
f(\T,\overline{\T})+\int d^2\theta \Sigma^3 P(\T)+ \int
d^2\overline{\theta}\, \overline{\Sigma}^3 \overline{
P}(\overline{\T}), \label{cc}
\end{equation}
where the chiral superfield $\Sigma$ is the conformal compensator,
\begin{equation}
\Sigma=1+\theta^2 M,
\end{equation}
and we use $\T$ to denote the radion superfield as well as its
lowest component. The function $f(\T,\overline{\T})$ is related to
the K\"ahler potential by
\begin{equation}
K(\T,\overline{\T})=-3\log \Big[-\frac {f(\T,\overline{\T})}
{3}\Big].
\end{equation}

To leading order in $1/L^2$, the warp factor $F(\phi) = e^{-k r_0
|\phi|}$. The conformal compensator is then
\begin{eqnarray}
\Sigma=\Sigma_0 e^{-k \T |\phi|},
\end{eqnarray}
where $\Sigma_0 =1+\theta^2 M$. The four-dimensional effective
lagrangian is obtained by integrating the conformal compensator
over the fifth dimension.   The kinetic term is
\begin{equation}
\frac k 2\int_{-\pi}^{\pi}
d^4\theta\,d\phi\,\Sigma_0\overline{\Sigma}_0~(\T+\overline{\T})e^{-k
(\T+\overline{\T})|\phi|}=\int d^4\theta ~\Sigma_0
\overline{\Sigma}_0\,\big(1-e^{-\pi k (\T+\overline{\T})}\big).
\end{equation}
This gives the same K\"ahler potential as found in the body of
this paper. The superpotential can be written as the sum of two
brane-localized constants,
\begin{equation}
\int_{-\pi}^\pi d^2\theta\,d\phi \Sigma^3_0\, e^{-3 k \T
|\phi|}\big[p_1~\delta(\phi)+p_2~ \delta(\phi-\pi)\big]=\int
d^2\theta \Sigma_0^3 \big[p_1+p_2~e^{-3 k \pi \T}\big].
\end{equation}
In each case, the correct radion dependence is given by the
conformal compensator.


\end{document}